\documentclass[twocolumn,pre,showpacs]{revtex4}
\usepackage{amssymb}
\usepackage{amsfonts}
\usepackage{amsmath}
\usepackage{graphicx}

\begin{document}

\title{Extended $q$-Gaussian and $q$-exponential distributions from Gamma
random variables}
\author{Adri\'{a}n A. Budini}
\affiliation{Consejo Nacional de Investigaciones Cient\'{\i}ficas y T\'{e}cnicas
(CONICET), Centro At\'{o}mico Bariloche, Avenida E. Bustillo Km 9.5, (8400)
Bariloche, Argentina, and Universidad Tecnol\'{o}gica Nacional (UTN-FRBA),
Fanny Newbery 111, (8400) Bariloche, Argentina}
\date{\today}

\begin{abstract}
The family of $q$-Gaussian and $q$-exponential probability densities fit the
statistical behavior of diverse complex self-similar non-equilibrium
systems. These distributions, independently of the underlying dynamics, can
rigorously be obtained by maximizing Tsallis \textquotedblleft
non-extensive\textquotedblright\ entropy under appropriate constraints, as
well as from superstatistical models. In this paper we provide an
alternative and complementary scheme for deriving these objects. We show
that $q$-Gaussian and $q$-exponential random variables can always be
expressed as function of two statistically independent Gamma random
variables with the same scale parameter. Their shape index determine the
complexity $q$-parameter. This result also allows to define an extended
family of asymmetric $q$-Gaussian and modified $q$-exponential densities,
which reduce to the previous ones when the shape parameters are the same.
Furthermore, we demonstrate that simple change of variables always allow to
relate any of these distributions with a Beta stochastic variable. The
extended distributions are applied in the statistical description of
different complex dynamics such as log-return signals in financial markets
and motion of point defects in fluid flows.
\end{abstract}

\maketitle


\section{Introduction}

Long-range interparticle interaction, long-term microscopic or mesoscopy
memory, fractal or multifractal occupation in phase space, cascade transfer
of energy or information, and intrinsic fluctuations of some dynamical
system parameters are some of the properties that nowadays are related with
complexity. One of the emergent properties related with these phenomena is
the power-law statistics of the corresponding nonequilibrium states. While
there exits different underlying formalism for tackling these issues,
maximization of \textit{Tsallis \textquotedblleft
non-extensive\textquotedblright\ entropy} \cite%
{tsallis,TsallisBook,Qlectures,qConstraints} provides an alternative basis
over which complexity can be analyzed and studied in a broad class of
systems. Introducing a generalized second moment constraint \cite%
{TsallisBook,qConstraints}, the formalism lead to a generalization of
standard normal probability densities, named as $q$-Gaussian distributions.
In terms of a generalized exponential function, they can be written as $%
P(x)=(\sqrt{\beta }/\mathcal{N}_{q})\exp _{q}(-\beta x^{2}),$ where the
parameter $q\in \mathrm{Re}$ defines different complexity classes. Explicitly, 
these statistical objects read%
\begin{equation}
P(x)=\frac{\sqrt{\beta }}{\mathcal{N}_{q}}\Big{[}1-(1-q)\beta x^{2}\Big{]}^{%
\frac{1}{1-q}},\ \ \ \ \ \ \ -\infty <q<1,  \label{qGaussLess}
\end{equation}%
where the variable of interest $x$ is restricted to the domain $0\leq
(1-q)\beta x^{2}\leq 1.$ On the other hand,%
\begin{equation}
P(x)=\frac{\sqrt{\beta }}{\mathcal{N}_{q}}\Big{[}\frac{1}{1+(q-1)\beta x^{2}}%
\Big{]}^{\frac{1}{q-1}},\ \ \ \ \ \ \ 1<q<3,  \label{qGaussMore}
\end{equation}%
where now $x$ is allowed to runs over the real line. The restriction $q<3$
follows from the normalization condition $\int_{-\infty }^{+\infty
}dxP(x)=1, $ which is guaranteed by the dimensionless constant $\mathcal{N}%
_{q}.$ The parameter $\sqrt{\beta }$ measures the width of the
distributions. As is well known \cite{TsallisBook}, in the limit $%
q\rightarrow 1$ both expressions reduce to the standard Gaussian
distribution. These generalizations allow to describe variables restricted
to a finite domain [Eq. (\ref{qGaussLess})] as well as power-law statistics
[Eq. (\ref{qGaussMore})].

$q$-Gaussian distributions also arise as solution of non-linear
Focker-Planck equations \cite{QFokker,LevyAsQ} as well as in the formulation
of central limit theorems with highly correlated random variables \cite%
{QCentralLimit}. Furthermore they fulfill a generalized fluctuation relation
symmetry \cite{budini}. A wide and diverse class of systems obey their
statistics \cite{TsallisBook}, such as in fluids flows \cite%
{superstatistics,lewis,beck}, optical lattices \cite{lutz}, trapped ions
interacting with a classical gas \cite{voe}, in granular mixtures \cite%
{puglisi}, anomalous diffusion in dusty plasma \cite{liu} or cellular
aggregates \cite{sawada}, avalanches sizes in earthquakes \cite{latora}, in
astrophysical variables \cite{voros}, as well as in econophysics \cite%
{osorio,borland,ausloos,rosenfeld,fuentes}.

When introducing a first moment constraint, Tsallis entropy leads to a $q$%
-exponential distribution [Eqs. (\ref{qGaussLess}) and (\ref{qGaussMore})
under the replacement $x^{2}\rightarrow x$ with $-\infty <q<1$ and $1<q<2$\
respectively], which in the limit $q\rightarrow 1$ recovers the standard
exponential probability density of a positive random variable. These
objects, for example, allow to fit high energy collisions \cite{wloda},
quark matter statistics \cite{biro}, solar flares \cite{stella} and momentum
distributions of charged hadrons \cite{khacha}. $q$-exponential functions
also fit anomalous power-law dipolar relaxation \cite{brouers} as well as
spin-glass relaxation \cite{pappas}. More recently, a kind of generalized $q$%
-Gamma probability density was introduced for describing stock trading
volume flow in financial markets \cite{queiros,souza,cortines}.

It is remarkable that all quoted probability densities can also be obtained
from a \textit{superstatistical} modeling \cite{superstatistics}, where a
parameter of an underlying probability measure becomes a (positive) random
variable characterized by a Gamma distribution \cite%
{feller,johnson,kleiber,vanKampen}. This is the case of $q$-Gaussian
densities, where the underlying distribution is a normal one \cite%
{superstatistics}, while for $q$-exponential it is an exponential function 
\cite{wloda}. For generalized $q$-Gamma variables the underlying
distribution is a Gamma density while the random parameter is distributed
according to an inverse Gamma distribution \cite{queiros}.

The main goal of this paper is to present a complementary and alternative
scheme to those provided by entropy extremization and superstatistics. We
show that random variables described by any of the quoted families of $q$%
-distributions can be written as a function of two independent (positive)
random Gamma variables \cite{feller,johnson,kleiber,vanKampen}. Their scale
parameter is assumed the same, while their shape indexes determine the
complexity $q$-parameter. When the shape indexes are different, a class of
extended asymmetric $q$-Gaussian and modified $q$-exponential distributions
are obtained. Generation of $q$-distributed random numbers is
straightforward from these results \cite{IEEE,levyNum}. We also show that
simple transformation of variables allow relating any of the obtained
densities with a Beta distribution. Interestingly, this statistical function
has been applied to model a wide variety of problems arising in different
disciplines \cite{feller,johnson,kleiber}. On the other hand, a $q$-triplet 
\cite{constantino,burlaga} for the probabilities densities is obtained. The
usefulness of the extended distributions in the context of financial signals 
\cite{ausloos} and motion of point defects in fluid flows \cite{beck} is
demonstrated. These systems are characterized by highly asymmetric
distributions. This feature is absent in previous approaches, being
recovered by the present one.

The paper is organized as follows. In Sec. II we review the properties of
Gamma random variables and introduce the main assumption over which the
present scheme relies. In Sec. III asymmetric $q$-Gaussian distributions are
obtained, while Sec. IV is devoted to modified $q$-exponential densities. In
Sec. V the properties of the proposed scheme as well as its applications are
discussed. In Sec. VI the Conclusions are provided.

\section{Model}

A stochastic random variable $Y$ is Gamma distributed \cite%
{feller,johnson,kleiber,vanKampen} if its probability density is%
\begin{equation}
P(y)=y^{\alpha -1}\frac{e^{-y/\theta }}{\theta ^{\alpha }\Gamma (\alpha )},\
\ \ \ \ y>0,\ \ \ \ \ 0<\alpha <\infty ,  \label{singleGamma}
\end{equation}%
where $\Gamma (\alpha )$ is the Gamma function. This distribution is
characterized by the scale parameter $\theta >0$\ and its shape parameter $%
\alpha .$ In terms of these parameters its mean value reads $\langle
Y\rangle =\int_{0}^{\infty }yP(y)dy=\alpha \theta ,$ with variance $\mathrm{%
var}[Y]=\langle Y^{2}\rangle -\langle Y\rangle ^{2}=\alpha \theta ^{2}.$ The
underlying stochastic process that leads to this statistic involves a
cascade-like mechanism \cite{feller,johnson,kleiber,vanKampen}. In fact,
this property is evident from the Laplace transform $P(u)=[\theta
^{-1}/(u+\theta ^{-1})]^{\alpha }$ where $P(u)=\int_{0}^{\infty
}dtP(t)e^{-ut}.$ Hence, when $\alpha $ is natural it reduces to a
convolution of exponential functions, which can be read as a cascade of
consecutive random steps.

The present approach relies on two independent Gamma random variables $%
Y_{1}, $ $Y_{2}.$ Their joint probability density then read%
\begin{equation}
P(y_{1},y_{2})=y_{1}^{\alpha -1}y_{2}^{\alpha ^{\prime }-1}\frac{%
e^{-(y_{1}+y_{2})/\theta }}{\theta ^{\alpha +\alpha ^{\prime }}\Gamma
(\alpha )\Gamma (\alpha ^{\prime })}.  \label{jointGamma}
\end{equation}%
Here, we assumed that both scale parameters $\theta $ are the same, while $%
\alpha $ and $\alpha ^{\prime }$ are the shape parameters of $Y_{1}$ and $%
Y_{2}$ respectively. The main ingredient of the present scheme is the ansatz%
\begin{equation}
X=f(Y_{1},Y_{2}),  \label{efe}
\end{equation}%
where the new random variable $X,$ depending of the function $%
f(y_{1},y_{2}), $\ develops different statistics. We will show that a wide
class of $q$-distributions arise from non-linear functions, which in turn
are asymmetric in their arguments (see Sec. III and IV). Nevertheless, in
all cases they fulfill a very simple symmetry (see Sec. V).

The probability distribution of $X$ is completely determined by the joint
probability (\ref{jointGamma}) and the function $f(y_{1},y_{2}).$ In fact,
it follows from a elementary change of variables \cite{vanKampen}. For
closing the problem, we introduce an extra random variable $Z$ defined by
the addition%
\begin{equation}
Z=(Y_{1}+Y_{2}).  \label{Zeta}
\end{equation}%
Therefore, the joint probability of $X$ and $Z$ is given by%
\begin{equation}
P(x,z)=P(y_{1},y_{2})|\det (J)|,  \label{Conjunta}
\end{equation}%
where $J$ is the Jacobian matrix%
\begin{equation}
J=\left( 
\begin{array}{ccc}
\frac{\partial y_{1}}{\partial x} &  & \frac{\partial y_{1}}{\partial z} \\ 
&  &  \\ 
\frac{\partial y_{2}}{\partial x} &  & \frac{\partial y_{2}}{\partial z}%
\end{array}%
\right) .
\end{equation}%
The probability of each variable follows by partial integration%
\begin{equation}
P(x)=\int_{0}^{\infty }dzP(x,z),\ \ \ \ \ \ \ \ \ P(z)=\int_{-\infty
}^{\infty }dxP(x,z).
\end{equation}%
As $Z$ is defined by the addition of two independent Gamma variables with
the same scale factor, it follows 
\begin{equation}
P(z)=z^{\alpha +\alpha ^{\prime }-1}\frac{e^{-z/\theta }}{\theta ^{\alpha
+\alpha ^{\prime }}\Gamma (\alpha +\alpha ^{\prime })}.  \label{PZetal}
\end{equation}%
Hence, $Z$ is also a Gamma variable [$Z>0,$ see Eq. (\ref{singleGamma})]
where its shape index is $(\alpha +\alpha ^{\prime })$ \cite%
{feller,johnson,kleiber,vanKampen}.

\section{Asymmetric $q$-Gaussian distributions}

In order to motivate the election of the function $f(y_{1},y_{2})$ that lead
to $q$-Gaussian statistics we may think (in a roughly way) in a Brownian
particle that interact with a complex reservoir. $Y_{1}$ and $Y_{2}$ are the
(positive and negative) impulse moments induced by the bath fluctuations. In
addition, the complexity of the system-environment interaction is taken into
account by a system response function $M^{-1}(Y_{1},Y_{2})$ that depends on
both $Y_{1}$ and $Y_{2}.$ Therefore, this contribution can be read as a
random-dissipative-like mechanism. The particle fluctuation is finally
written as $X\approx (Y_{1}-Y_{2})/M(Y_{1},Y_{2}).$ One may also think in a
economical agent that, from the available information, predict that a future
price may increases or decrease a quantity $Y_{1}$ or $Y_{2}$ respectively.
The weight of the available information is then measured by $%
M^{-1}(Y_{1},Y_{2}),$ leading to the same kind of dependence.

In order to close the model, we assume that $M(Y_{1},Y_{2})$ is given by a
kind of average or mean value between the two random values $Y_{1}$ and $%
Y_{2}.$ Specifically we take%
\begin{equation}
M(Y_{1},Y_{2})=\Big{[}\frac{1}{2}(Y_{1}^{\mu }+Y_{2}^{\mu })\Big{]}^{1/\mu },
\label{PromedioGeneralizado}
\end{equation}
where $\mu \in \mathrm{Re}.$ Therefore, we write the $X$ random variable
[Eq. (\ref{efe})] as 
\begin{equation}
X=\frac{1}{\sqrt{\beta }}\frac{Y_{1}-Y_{2}}{2\Big{[}\frac{1}{2}(Y_{1}^{\mu
}+Y_{2}^{\mu })\Big{]}^{1/\mu }}.  \label{GenerlizedMoment}
\end{equation}%
By convenience we introduced a factor one half. On the other hand, the
additional parameter $\sqrt{\beta }>0$ scales and gives the right units of $%
X.$ In fact, notice that the remaining contribution in Eq. (\ref%
{GenerlizedMoment}) is dimensionless. Taking different values of the real
parameter $\mu $ a wide class of probability distributions arise, which in
turn may also depends on the parameters $\theta ,$\ $\alpha $ and $\alpha
^{\prime }$ that determine the joint probability density (\ref{jointGamma}).

\subsection{Arithmetic mean}

The arithmetic mean value corresponds to $\mu =1,$ implying that%
\begin{equation}
X=\frac{1}{\sqrt{\beta }}\frac{Y_{1}-Y_{2}}{Y_{1}+Y_{2}}.  \label{Aritmetico}
\end{equation}%
Notice that, for any possible value of $Y_{1}$ and $Y_{2},$ the random
variable $X$ assume bounded values in the domain $(-1/\sqrt{\beta },+1/\sqrt{%
\beta }).$ Taking into account the $Z$ variable [Eq. (\ref{Zeta})], we
obtain the following inverted relations%
\begin{equation}
Y_{1}=\frac{Z}{2}(1+\sqrt{\beta }X),\ \ \ \ \ \ \ Y_{2}=\frac{Z}{2}(1-\sqrt{%
\beta }X),
\end{equation}%
which in turn implies that $|\det (J)|=\sqrt{\beta }z/2.$ Eqs. (\ref%
{jointGamma}) and (\ref{Conjunta}) lead to $P(x,z)=P(x)P(z),$ where $P(z)$\
is given by Eq. (\ref{PZetal}). Therefore, the random variables $X$ and $Z$
are statistically independent. Furthermore, $X$ obeys the statistics given
by the probability density%
\begin{equation}
P(x)=\frac{\sqrt{\beta }}{\mathcal{N}_{\alpha \alpha ^{\prime }}}(1+\sqrt{%
\beta }x)^{\alpha -1}(1-\sqrt{\beta }x)^{\alpha ^{\prime }-1},
\label{PqLessAsimetrica}
\end{equation}%
where the normalization constant reads $\mathcal{N}_{\alpha \alpha ^{\prime
}}=2^{\alpha +\alpha ^{\prime }-1}\Gamma (\alpha )\Gamma (\alpha ^{\prime
})/\Gamma (\alpha +\alpha ^{\prime }).$ Notice that $P(x)$ does not depends
on the scale parameter $\theta $\ [see Eq. (\ref{jointGamma})]. It only
depends on the shape indexes $\alpha ,$ $\alpha ^{\prime },$ and the scale
parameter $\beta .$

The distribution (\ref{PqLessAsimetrica}) develop a maximum located at%
\begin{equation}
x_{M}=\frac{1}{\sqrt{\beta }}\frac{\alpha -\alpha ^{\prime }}{(\alpha
+\alpha ^{\prime }-2)},  \label{Xmqmenor}
\end{equation}%
when $\alpha >1,$ $\alpha ^{\prime }>1,$ or at $x_{M}=\pm 1/\sqrt{\beta }$
in any other case. Its average value read%
\begin{equation}
\langle X\rangle =\frac{1}{\sqrt{\beta }}\frac{\alpha -\alpha ^{\prime }}{%
\alpha +\alpha ^{\prime }},
\end{equation}%
while the variance $\mathrm{var}[X]=\langle X^{2}\rangle -\langle X\rangle
^{2}$ is given by%
\begin{equation}
\mathrm{var}[X]=\frac{1}{\beta }\frac{4\alpha \alpha ^{\prime }}{(\alpha
+\alpha ^{\prime })^{2}(1+\alpha +\alpha ^{\prime })}.
\end{equation}

\subsubsection*{$q$-Gaussian distributions}

Eq. (\ref{PqLessAsimetrica}) can be rewritten as%
\begin{equation}
P(x)=\frac{\sqrt{\beta }}{\mathcal{N}_{\alpha \alpha ^{\prime }}}(1-\beta
x^{2})^{\frac{\alpha +\alpha ^{\prime }}{2}-1}\Big{(}\frac{1+\sqrt{\beta }x}{%
1-\sqrt{\beta }x}\Big{)}^{\frac{\alpha -\alpha ^{\prime }}{2}}.
\label{AsymetricaLess}
\end{equation}%
Hence, we name this function as an asymmetric Poissonian $q$-Gaussian
distribution $G_{<1}^{p}(x|q,a,\beta )$ with index $q,$ and asymmetry
parameter $a,$%
\begin{equation}
q=1-\Big{[}\frac{\alpha +\alpha ^{\prime }}{2}-1\Big{]}^{-1},\ \ \ \ \ \ \ \
\ \ a=\frac{\alpha -\alpha ^{\prime }}{2}.  \label{qGaussMenorUno}
\end{equation}%
From the positivity of $\alpha $ and $\alpha ^{\prime }$ the asymmetry index
must to satisfy $|a|<(2-q)/(1-q).$ In the symmetric case, $a=0,$ $\alpha
^{\prime }=\alpha ,$ we get%
\begin{equation}
P(x)=\frac{\sqrt{\beta }}{\mathcal{N}_{\alpha }}(1-\beta x^{2})^{\alpha -1},
\end{equation}%
where $\mathcal{N}_{\alpha }=2^{2\alpha -1}\Gamma ^{2}(\alpha )/\Gamma
(2\alpha ).$ Therefore, under the association $\beta \rightarrow \beta
(1-q), $ with $\alpha >1,$ we recover Eq. (\ref{qGaussLess}). Over the
domain $\alpha \in (1,\infty ),$ the non-extensive parameter runs in the
interval $q\in (-\infty ,1).$

In Fig. 1 we plot the function (\ref{AsymetricaLess}) for different values
of the asymmetric factor $a,$ Eq. (\ref{qGaussMenorUno}). For increasing $%
a>0,$ the distribution accumulates around $\sqrt{\beta }x\approx 1.$ For $%
a<0 $ a reflected accumulation around $\sqrt{\beta }x\approx -1$ is
developed. The distribution with $a=0$ corresponds to Tsallis non-extensive
thermodynamics. 
\begin{figure}[tbp]
\includegraphics[bb=18 284 420 595,width=8.75cm]{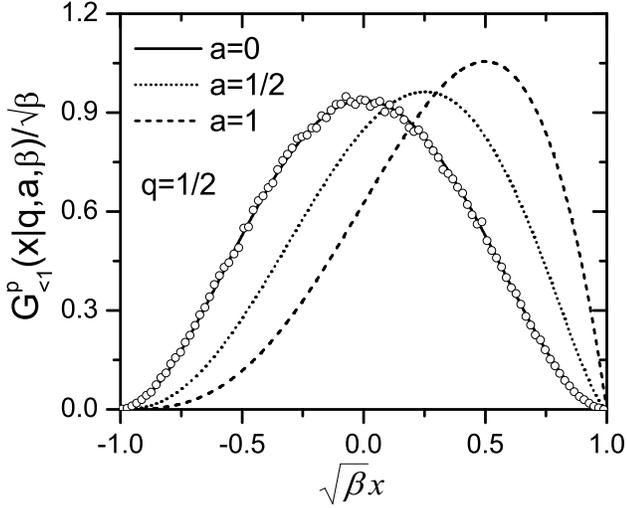}
\caption{Poissonian $q$-Gaussian probability distribution Eq. (\protect\ref%
{AsymetricaLess}), for different values of the asymmetry $a,$ Eq. (\protect
\ref{qGaussMenorUno}). The circles correspond to a numerical simulation
based on Eq. (\protect\ref{Aritmetico}).}
\end{figure}

\subsection{Geometric mean}

In Eq. (\ref{PromedioGeneralizado}) the geometric mean correspond to $\lim
\mu \rightarrow 0,$ which satisfies $\lim_{\mu \rightarrow 0}\Big{[}\frac{1}{%
2}(Y_{1}^{\mu }+Y_{2}^{\mu })\Big{]}^{1/\mu }=\sqrt{Y_{1}Y_{2}}.$ Therefore,
we get the random variable [Eq. (\ref{GenerlizedMoment})]%
\begin{equation}
X=\frac{1}{\sqrt{\beta }}\frac{Y_{1}-Y_{2}}{2\sqrt{Y_{1}Y_{2}}}.
\label{Geometrico}
\end{equation}
Notice that $X$ assumes values over the entire real number line, $X\in 
\mathrm{Re}.$ In this case, the inverted relations are%
\begin{equation}
Y_{1}=\frac{Z}{2}\Big{(}1+\frac{\sqrt{\beta }X}{\sqrt{1+\beta X^{2}}}\Big{)}%
,\ \ \ \ \ \ \ Y_{2}=\frac{Z}{2}\Big{(}1-\frac{\sqrt{\beta }X}{\sqrt{1+\beta
X^{2}}}\Big{)},
\end{equation}%
implying that $|\det (J)|=\sqrt{\beta }(z/2)(1+\beta x^{2})^{-3/2},$ which
in turn also lead to $P(x,z)=P(x)P(z),$ but here%
\begin{eqnarray}
P(x) &=&\frac{\sqrt{\beta }}{\mathcal{N}_{\alpha \alpha ^{\prime }}}\Big{(}%
\frac{1}{1+\beta x^{2}}\Big{)}^{\frac{\alpha +\alpha ^{\prime }+1}{2}}(\sqrt{%
1+\beta x^{2}}+\sqrt{\beta }x)^{\alpha -1}  \notag \\
&&(\sqrt{1+\beta x^{2}}-\sqrt{\beta }x)^{\alpha ^{\prime }-1}.
\label{PqMoreAsimetrica}
\end{eqnarray}%
As in the previous case, this distribution is independent of the rate
parameter $\theta .$\ The normalization constant is the same, $\mathcal{N}%
_{\alpha \alpha ^{\prime }}=2^{\alpha +\alpha ^{\prime }-1}\Gamma (\alpha
)\Gamma (\alpha ^{\prime })/\Gamma (\alpha +\alpha ^{\prime }).$

Eq. (\ref{PqMoreAsimetrica}) develops a maximum, which occurs at%
\begin{equation}
x_{M}=\frac{\alpha -\alpha ^{\prime }}{\sqrt{(1+2\alpha )(1+2\alpha ^{\prime
})\beta }}.  \label{xm}
\end{equation}%
In the limit $\sqrt{\beta }x\gg 1,$ a power-law behavior arise%
\begin{equation}
\lim_{x\rightarrow \infty }P(x)\approx \frac{\sqrt{\beta }}{\mathcal{N}%
_{\alpha \alpha ^{\prime }}}2^{\alpha -\alpha ^{\prime }}\Big{(}\frac{1}{%
\sqrt{\beta }x}\Big{)}^{2\alpha ^{\prime }+1},  \label{powerLawPositivo}
\end{equation}%
while for $\sqrt{\beta }x\ll -1$ we obtain%
\begin{equation}
\lim_{x\rightarrow -\infty }P(x)\approx \frac{\sqrt{\beta }}{\mathcal{N}%
_{\alpha \alpha ^{\prime }}}2^{\alpha ^{\prime }-\alpha }\Big{(}\frac{1}{-%
\sqrt{\beta }x}\Big{)}^{2\alpha +1}.  \label{powerLawNegativo}
\end{equation}%
Due to the previous asymptotic behaviors the moments are not defined for any
value of the characteristic shape parameters. When $\alpha >1/2$ and $\alpha
^{\prime }>1/2,$ the average value reads%
\begin{equation}
\langle X\rangle =\frac{1}{\sqrt{\beta }}(\alpha -\alpha ^{\prime })\frac{%
\Gamma (\alpha -\frac{1}{2})\Gamma (\alpha ^{\prime }-\frac{1}{2})}{2\Gamma
(\alpha )\Gamma (\alpha ^{\prime })},  \label{X}
\end{equation}%
while the second moment, for $\alpha >1$ and $\alpha ^{\prime }>1,$ is%
\begin{equation}
\langle X^{2}\rangle \!=\!\frac{1}{\beta }[(\alpha -\alpha ^{\prime
})^{2}+(\alpha +\alpha ^{\prime }-2)]\frac{\Gamma (\alpha -1)\Gamma (\alpha
^{\prime }-1)}{4\Gamma (\alpha )\Gamma (\alpha ^{\prime })}.  \label{X^2}
\end{equation}%
Outside the previous intervals the first two moments are not defined. 
\begin{figure}[tbp]
\includegraphics[bb=18 284 420 595,width=8.75cm]{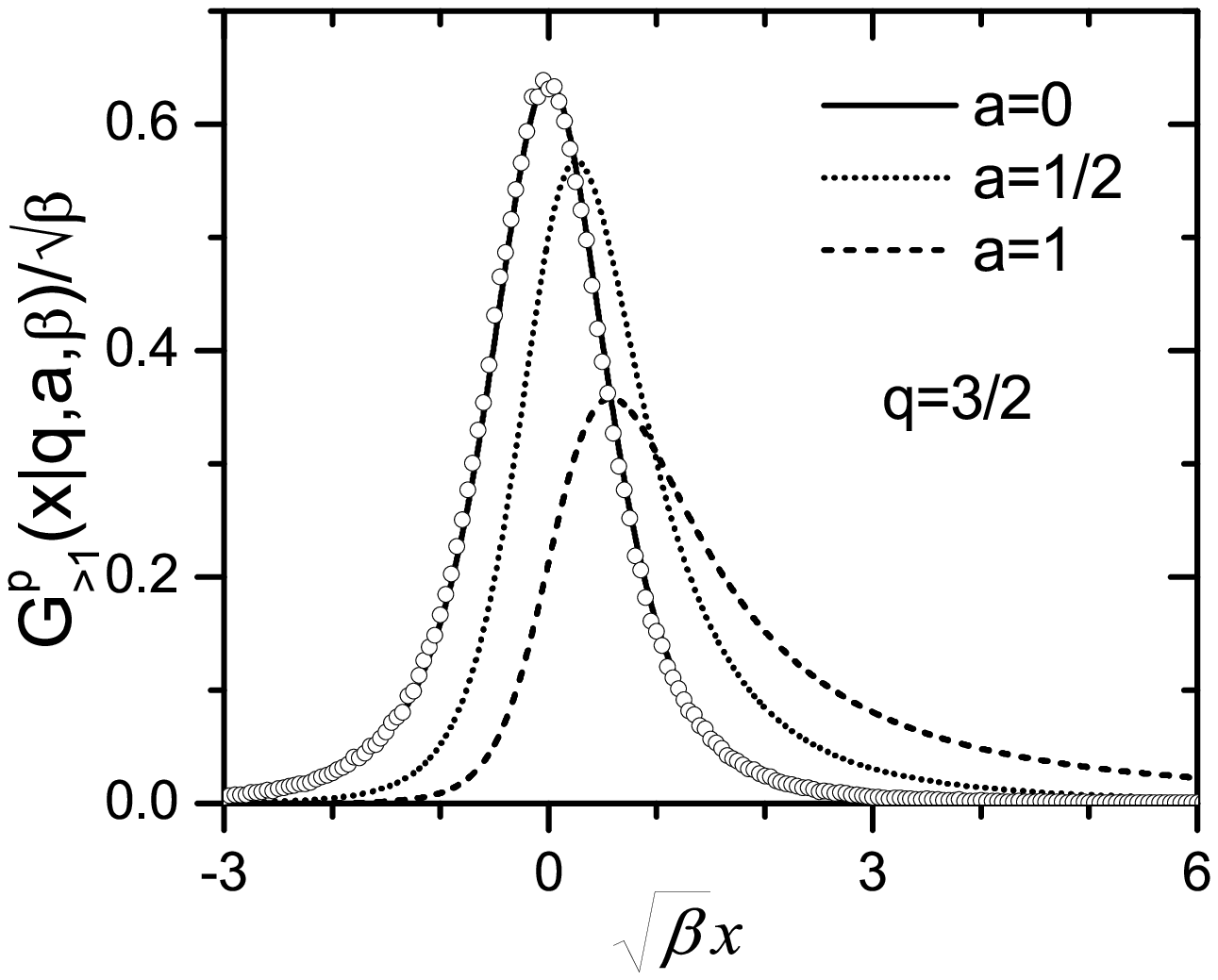}
\caption{Poissonian $q$-Gaussian probability distribution Eq. (\protect\ref%
{AsymetricaMore}), for different values of the asymmetry $a,$ Eq. (\protect
\ref{qGaussMayorUno}). The circles correspond to a numerical simulation
based on Eq. (\protect\ref{Geometrico}).}
\end{figure}

\subsubsection*{$q$-Gaussian distributions}

Eq. (\ref{PqMoreAsimetrica}) can be rewritten as%
\begin{equation}
P(x)=\frac{\sqrt{\beta }}{\mathcal{N}_{\alpha \alpha ^{\prime }}}\Big{(}%
\frac{1}{1+\beta x^{2}}\Big{)}^{\frac{\alpha +\alpha ^{\prime }+1}{2}}\Big{(}%
\frac{\sqrt{1+\beta x^{2}}+\sqrt{\beta }x}{\sqrt{1+\beta x^{2}}-\sqrt{\beta }%
x}\Big{)}^{\frac{\alpha -\alpha ^{\prime }}{2}}.  \label{AsymetricaMore}
\end{equation}%
As is the previous case, we name this function as an asymmetric Poissonian $%
q $-Gaussian distribution $G_{>1}^{p}(x|q,a,\beta )$ with index $q,$ and
asymmetry parameter $a,$%
\begin{equation}
q=1+\Big{[}\frac{\alpha +\alpha ^{\prime }}{2}+\frac{1}{2}\Big{]}^{-1},\ \ \
\ \ \ \ \ \ \ a=\frac{\alpha -\alpha ^{\prime }}{2}.  \label{qGaussMayorUno}
\end{equation}%
Hence, $1<q<3$ and the asymmetry factor satisfy the restriction $%
|a|<(3-q)/[2(q-1)].$ In the symmetric case, $a=0,$ $\alpha ^{\prime }=\alpha
,$ $P(x)$ reduces to%
\begin{equation}
P(x)=\frac{\sqrt{\beta }}{\mathcal{N}_{\alpha }}\Big{(}\frac{1}{1+\beta x^{2}%
}\Big{)}^{\alpha +\frac{1}{2}},
\end{equation}%
where $\mathcal{N}_{\alpha }=2^{2\alpha -1}\Gamma ^{2}(\alpha )/\Gamma
(2\alpha ).$ Under the association $\beta \rightarrow \beta (q-1),$ we
recover Eq. (\ref{qGaussMore}). In the interval $\alpha \in (0,\infty )$ the
non-extensive parameter runs in the interval $q\in (3,1).$

In Fig. 2 we plot the function (\ref{AsymetricaMore}) for different values
of the asymmetric factor $a,$ Eq. (\ref{qGaussMayorUno}). For increasing $%
a>0,$ the maximum of the distribution is shifted to higher values. For $a<0$
the extremum develops for negative values. The symmetric case $a=0$
corresponds to the $q$-Gaussian distribution arising from Tsallis entropy.

\subsection{Relation between both cases}

Given $X$ determinate by relation (\ref{Aritmetico}), the random variable $%
X^{\prime }$ defined as%
\begin{equation}
X^{\prime }=\frac{X}{\sqrt{1-\beta X^{2}}},
\end{equation}%
recover Eq. (\ref{Geometrico}). This simple relation demonstrate that there
exist a one to one mapping between $q$-Gaussian variables in the different
domains of the complexity parameter $q.$ In fact, if we define$\sqrt{\beta }%
X=\sin (\phi )\in (-1,1)$ for $q\in (-\infty ,1),$ hence $\sqrt{\beta }%
X^{\prime }=\tan (\phi )\in (-\infty ,+\infty ),$ where $X^{\prime }$ has
associated the index $q\in (1,3).$

Alternatively, if $X$ is given by Eq. (\ref{Geometrico}), the inverse
transformation%
\begin{equation}
X^{\prime }=\frac{X}{\sqrt{1+\beta X^{2}}},
\end{equation}%
lead to Eq. (\ref{Aritmetico}). While these relations are known for
symmetric $q$-Gaussian distributions $(a=0)$\ \cite{TsallisBook}, here we
showed that they are also valid for the asymmetric densities $(a\neq 0)$
introduced previously.

\section{Modified $q$-exponential distributions}

Variables distributed according to a $q$-exponential density are positive.
Therefore, the previous scheme does not apply, but a similar one can be
implemented. We name the emerging distributions as modified $q$-exponential
densities. Taking into account the notation of Refs. \cite%
{queiros,souza,cortines} they can also be called as generalized $q$-Gamma
densities. Nevertheless, it seems that they do not satisfy the same entropic
properties than standard Gamma distributions \cite{stolongo}. On the other
hand, we remark that some properties of the following distributions are
known and can be found under different denominations \cite{johnson,kleiber}.

\subsection{Bounded domain}

For getting a positive variable, we introduce the following functional
dependence%
\begin{equation}
X=\frac{1}{\sqrt{\beta }}\frac{Y_{2}}{Y_{1}+Y_{2}}.  \label{XExpoLess}
\end{equation}%
Notice that this assumption is very similar to Eq. (\ref{Aritmetico}), but
here $X$ is a bounded positive stochastic variable, $0<\sqrt{\beta }X<1.$
Using the approach defined in Sec. II, here we obtain the inverse relations%
\begin{equation}
Y_{1}=Z(1-\sqrt{\beta }X),\ \ \ \ \ \ \ \ \ \ \ Y_{2}=Z\sqrt{\beta }X,
\end{equation}%
while $|\det (J)|=\sqrt{\beta }z,$ leading again to a statistical
independence of $X$ and $Z,$ that is $P(x,z)=P(x)P(z).$ The density of
interest here is%
\begin{equation}
P(x)=\frac{\sqrt{\beta }}{\mathcal{N}_{\alpha \alpha ^{\prime }}}(\sqrt{%
\beta }x)^{\alpha ^{\prime }-1}(1-\sqrt{\beta }x)^{\alpha -1},
\label{DistortedExpQless}
\end{equation}%
where $\mathcal{N}_{\alpha \alpha ^{\prime }}=\Gamma (\alpha )\Gamma (\alpha
^{\prime })/\Gamma (\alpha +\alpha ^{\prime }).$ When $\alpha >1$ and $%
\alpha ^{\prime }>1,$ $P(x)$ reaches a maximal value located at%
\begin{equation}
x_{M}=\frac{1}{\sqrt{\beta }}\frac{\alpha ^{\prime }-1}{(\alpha +\alpha
^{\prime }-2)},
\end{equation}%
Its first moment reads%
\begin{equation}
\langle X\rangle =\frac{1}{\sqrt{\beta }}\frac{\alpha ^{\prime }}{\alpha
+\alpha ^{\prime }},
\end{equation}%
while the variance is given by%
\begin{equation}
\mathrm{var}[X]=\frac{1}{\beta }\frac{\alpha \alpha ^{\prime }}{(\alpha
+\alpha ^{\prime })^{2}(1+\alpha +\alpha ^{\prime })}.
\end{equation}

\subsubsection*{$q$-exponential densities}

The distribution (\ref{DistortedExpQless}) may be named as a modified
Poissonian $q$-exponential distribution $E_{<1}^{p}(x|q,d,\beta )$ with
index $q,$ and \textquotedblleft distortion parameter\textquotedblright\ $d,$%
\begin{equation}
q=1-\frac{1}{\alpha -1},\ \ \ \ \ \ \ \ \ \ d=\alpha ^{\prime }-1.
\label{qExpoMenorUno}
\end{equation}%
Therefore, $-\infty <q<1$ and $d>-1.$ When $d=0,$ that is $\alpha ^{\prime
}=1,$ Eq. (\ref{DistortedExpQless}) becomes%
\begin{equation}
P(x)=\frac{\sqrt{\beta }}{\mathcal{N}_{\alpha }}(1-\sqrt{\beta }x)^{\alpha
-1},
\end{equation}%
where $\mathcal{N}_{\alpha }=\Gamma (\alpha )/\Gamma (\alpha +1).$
Therefore, under the extra association $\sqrt{\beta }\rightarrow \sqrt{\beta 
}(1-q),$ we obtain a standard $q$-exponential density. For $\alpha \in
(1,\infty ),$ it follows $q\in (-\infty ,1).$

In Fig. 3 we plotted the function (\ref{DistortedExpQless}) for different
values of the distortion parameter $d,$ Eq. (\ref{qExpoMenorUno}). For $d<0,$
the density diverges around the origin. This property is inherited from the
Gamma distribution Eq. (\ref{jointGamma}). On the other hand, for $d>0$ the
density vanishes at the origin and for increasing $d$ it accumulates around $%
\sqrt{\beta }x\approx 1.$ The plot for $a=0$ is the $q$-exponential
distribution arising from Tsallis entropy. 
\begin{figure}[tbp]
\includegraphics[bb=18 284 420 595,width=8.75cm]{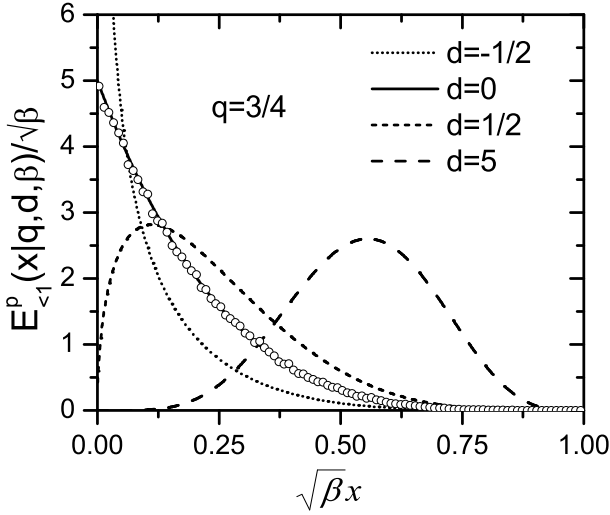}
\caption{Poissonian $q$-exponential probability distribution Eq. (\protect
\ref{DistortedExpQless}), for different values of the distortion parameter $%
d,$ Eq. (\protect\ref{qExpoMenorUno}). The circles correspond to a numerical
simulation based on Eq. (\protect\ref{XExpoLess}).}
\end{figure}

\subsection{Unbounded domain}

An unbounded positive variable $(0<X<\infty )$ is obtained from the relation%
\begin{equation}
X=\frac{1}{\sqrt{\beta }}\frac{Y_{2}}{Y_{1}},  \label{Cociente}
\end{equation}%
where as in the previous cases $\sqrt{\beta }$ scales the random variable $%
X. $ Here, the inverse relations are%
\begin{equation}
Y_{1}=\frac{Z}{1+\sqrt{\beta }X},\ \ \ \ \ \ \ \ \ Y_{2}=\frac{Z\sqrt{\beta }%
X}{1+\sqrt{\beta }X},
\end{equation}%
while $|\det (J)|=\sqrt{\beta }z/(1+\sqrt{\beta }x)^{2},$ leading to $%
P(x,z)=P(x)P(z),$ where%
\begin{equation}
P(x)=\frac{\sqrt{\beta }}{\mathcal{N}_{\alpha \alpha ^{\prime }}}\frac{(%
\sqrt{\beta }x)^{\alpha ^{\prime }-1}}{(1+\sqrt{\beta }x)^{\alpha +\alpha
^{\prime }}},  \label{DistortedUp}
\end{equation}%
with $\mathcal{N}_{\alpha \alpha ^{\prime }}=\Gamma (\alpha )\Gamma (\alpha
^{\prime })/\Gamma (\alpha +\alpha ^{\prime }).$ This distribution develops
a maximum that is located at $(\alpha ^{\prime }>1)$%
\begin{equation}
x_{M}=\frac{1}{\sqrt{\beta }}\frac{\alpha ^{\prime }-1}{(\alpha +1)},
\end{equation}%
For$\sqrt{\beta }x\gg 1,$ it follows the asymptotic power-law behavior%
\begin{equation}
\lim_{x\rightarrow \infty }P(x)\approx \frac{\sqrt{\beta }}{\mathcal{N}%
_{\alpha \alpha ^{\prime }}}\Big{(}\frac{1}{\sqrt{\beta }x}\Big{)}^{\alpha
+1}.
\end{equation}%
In consequence, the moments are not defined for any value of the shape
parameter $\alpha .$ For $\alpha >1,$ the mean value reads%
\begin{equation}
\langle X\rangle =\frac{1}{\sqrt{\beta }}\frac{\Gamma (\alpha -1)\Gamma
(\alpha ^{\prime }+1)}{\Gamma (\alpha )\Gamma (\alpha ^{\prime })},
\end{equation}%
while the second moment is only defined for $\alpha >2,$%
\begin{equation}
\langle X^{2}\rangle =\frac{1}{\beta }\frac{\Gamma (\alpha -2)\Gamma (\alpha
^{\prime }+2)}{\Gamma (\alpha )\Gamma (\alpha ^{\prime })}.
\end{equation}

\subsubsection*{$q$-exponential densities}

The distribution (\ref{DistortedExpQless}) may also be named as a modified
Poissonian $q$-exponential distribution $E_{>1}^{p}(x|q,d,\beta )$ with
index $q,$ and distortion parameter $d,$%
\begin{equation}
q=1+\frac{1}{\alpha +\alpha ^{\prime }},\ \ \ \ \ \ \ \ \ \ d=\alpha
^{\prime }-1.  \label{qExpoMayorUno}
\end{equation}%
In consequence, $1<q<\infty ,$ and the distortion parameter satisfy $%
(2-q)/(q-1)>d>-1.$ When the distortion is null, $\alpha ^{\prime }=1,$ Eq. (%
\ref{DistortedUp}) becomes%
\begin{equation}
P(x)=\frac{\sqrt{\beta }}{\mathcal{N}_{\alpha }}\frac{1}{(1+\sqrt{\beta }%
x)^{\alpha +1}},
\end{equation}%
where $\mathcal{N}_{\alpha }=\Gamma (\alpha )/\Gamma (1+\alpha ).$ Thus,
under the extra association $\sqrt{\beta }\rightarrow \sqrt{\beta }(1-q)$ we
get a $q$-exponential probability density. In this case, for $\alpha \in
(0,\infty ),$ it follows $q\in (2,1).$

The function (\ref{DistortedUp}) is plotted in Fig. 4 for different values
of the distortion parameter $d,$ Eq. (\ref{qExpoMayorUno}). For $d<0,$ the
density diverges around the origin. On the other hand, for $d>0$ the density
vanishes at the origin. In all cases a power-law behavior is obtained for $%
\sqrt{\beta }x\gg 1.$ The plot for $d=0$ is the $q$-exponential distribution
arising from Tsallis entropy. 
\begin{figure}[tbp]
\includegraphics[bb=18 284 420 595,width=8.75cm]{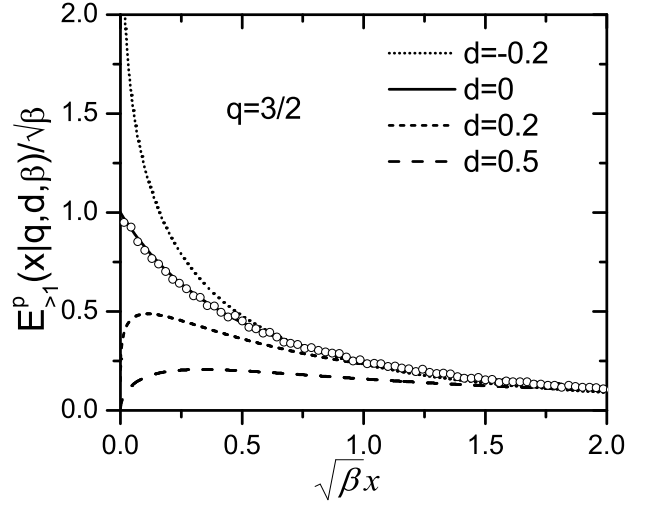}
\caption{Poissonian $q$-exponential probability distribution Eq. (\protect
\ref{DistortedUp}), for different values of the distortion parameter $d,$
Eq. (\protect\ref{qExpoMayorUno}). The circles correspond to a numerical
simulation based on Eq. (\protect\ref{Cociente}).}
\end{figure}

\subsection{Relation between both cases}

Given $X$ determinate by Eq. (\ref{XExpoLess}), the random variable $%
X^{\prime }$ defined as%
\begin{equation}
X^{\prime }=\frac{X}{1-\sqrt{\beta }X},
\end{equation}%
is given by Eq. (\ref{Cociente}). Alternatively, if $X$ is given by Eq. (\ref%
{Cociente}), the inverse transformation%
\begin{equation}
X^{\prime }=\frac{X}{1+\sqrt{\beta }X},
\end{equation}%
lead to Eq. (\ref{XExpoLess}). These conjugate relations are valid for both
the unmodified $(d=0)$ as well as the modified $(d\neq 0)$ $q$-exponential
densities.

\subsection{Stretched $q$-exponential densities}

Introducing the change of variables%
\begin{equation}
\sqrt{\tilde{\beta}}\tilde{X}=\Big{(}\sqrt{\beta }X\Big{)}^{1/\nu },
\label{estrecho}
\end{equation}
defined by the extra parameter $\nu \in \mathrm{Re},$\ if $X$ is distributed
according to Eq. (\ref{DistortedExpQless}), it follows $[P(\tilde{x})d\tilde{%
x}=P(x)dx]$%
\begin{equation}
P(\tilde{x})=\frac{\nu }{\mathcal{N}_{\alpha \alpha ^{\prime }}}\Big{(}\sqrt{%
\tilde{\beta}}\tilde{x}\Big{)}^{\nu \alpha ^{\prime }-1}\Big{[}1-\Big{(}%
\sqrt{\tilde{\beta}}\tilde{x}\Big{)}^{\nu }\Big{]}^{\alpha -1}.
\label{estrechadaMenorUno}
\end{equation}%
On other hand, if $X$ is distributed according to Eq. (\ref{DistortedUp}),
we get%
\begin{equation}
P(\tilde{x})=\frac{\nu \sqrt{\tilde{\beta}}}{\mathcal{N}_{\alpha \alpha
^{\prime }}}\frac{\Big{(}\sqrt{\tilde{\beta}}\tilde{x}\Big{)}^{\nu \alpha
^{\prime }-1}}{\Big{[}1+(\sqrt{\tilde{\beta}}\tilde{x})^{\nu }\Big{]}%
^{\alpha +\alpha ^{\prime }}}.  \label{StretchedPower}
\end{equation}%
In both cases, imposing the condition $\nu \alpha ^{\prime }=1,$ the
previous two expressions becomes stretched $q$-exponential densities, $P(x)=(%
\sqrt{\beta }/\mathcal{N}_{q})\exp _{q}[-(\sqrt{\beta }x)^{\nu }]$ $(x>0,$ $%
\nu >0).$ Hence, these distributions can also be covered with the present
approach [Eqs. (\ref{XExpoLess}) and (\ref{Cociente}) under the change of
variables (\ref{estrecho})].

\section{Properties and applications}

In the previous two sections we demonstrated that the assumption (\ref{efe})
allow us to recover and to define an extend family of $q$-Gaussian and $q$%
-exponential densities. Here, we discuss some general properties of the
approach as well as some applications of the extended distributions.

\subsection{$q$-distributed random numbers}

Numerical generation of random numbers obeying $q$-Gaussian statistics was
explored previously by introducing a generalized Box-Muller method \cite%
{IEEE}. Generation of Levy distributed numbers was also established \cite%
{levyNum}. On the other hand, numerical generation of Gamma random numbers
is also well established \cite{johnson,kleiber}. Therefore, the present
scheme defines an alternative and solid basis for obtaining $q$-distributed
random numbers by generating two independent Gamma random numbers. Using
this method, in Fig. (1) to (4) we explicitly show (circles) the recovering
of the symmetric and unmodified distributions, all of then corresponding to
Tsallis entropy formalism.

\subsection{Symmetries}

While the underlying joint statistics of the Gamma variables depends on the
scale parameter $\theta ,$\ Eq. (\ref{jointGamma}), the distributions of $X$
do not depend on it. This is not the only symmetry of the proposed scheme.
In fact, it is simple to check that the symmetry%
\begin{equation}
f(Y_{1},Y_{2})=f(\frac{1}{Y_{2}},\frac{1}{Y_{1}}),  \label{symetria}
\end{equation}%
is fulfilled, where $f(Y_{1},Y_{2})$ define the $X$ random variable, Eq. (%
\ref{efe}). In fact, this property is valid for the $q$-Gaussian case [Eq. (%
\ref{GenerlizedMoment}) for any $\mu $] as well as for the $q$-exponential
variables [Eqs. (\ref{XExpoLess}) and (\ref{Cociente})]. We notice that $%
f(Y_{1},Y_{2})=g(Y_{1}/Y_{2}),$ for arbitrary functions $g(y),$ always
satisfies the relation (\ref{symetria}). Extra structures can be established
by introducing arbitrary change of variables.

The symmetry (\ref{symetria}) implies that the same results arise if instead
of Gamma distributed variables one take inverse Gamma variables, that is, $%
Y^{\prime }=1/Y$\ where $Y$ is Gamma distributed [Eq. (\ref{singleGamma})].
Using that $P(y)dy=P(y^{\prime })dy^{\prime },$ it follows%
\begin{equation}
P(y^{\prime })=\frac{1}{(y^{\prime })^{\alpha +1}}\frac{e^{-1/y^{\prime
}\theta }}{\theta ^{\alpha }\Gamma (\alpha )},\ \ \ \ \ \ \ 0<\alpha <\infty.
\end{equation}

\subsection{Relation with Beta distributions}

As all functions $f(Y_{1},Y_{2})$ fulfill the condition (\ref{symetria}), it
is clear that any of the corresponding variables $X$ are always related by a
change of variables between them. Therefore, it does not make sense to
affirm that one of them generates or is more fundamental than the others.
Nevertheless, here we want to emphasize that any of the probability
densities obtained previously can be related with the well known Beta
distribution \cite{feller,johnson,kleiber}. It reads%
\begin{equation}
P(w)=\frac{\Gamma (\alpha +\alpha ^{\prime })}{\Gamma (\alpha )\Gamma
(\alpha ^{\prime })}w^{\alpha -1}(1-w)^{\alpha ^{\prime }-1},  \label{Beta}
\end{equation}%
where the domain of its variable is $w\in (0,1).$ Furthermore, its shape
parameters $\alpha $ and $\alpha ^{\prime }$\ are positive.

Defining the change of variables $w=w(x),$ all obtained $q$-distributions
becomes equal to Eq. (\ref{Beta}). Alternatively, defining a new variable $%
x=x(w),$ from the Beta distribution it is possible to obtain the $q$%
-densities. Explicitly, for the asymmetric distribution Eq. (\ref%
{PqLessAsimetrica}) [or Eq. (\ref{AsymetricaLess})], the change of variables
read%
\begin{equation}
w=\frac{1}{2}(1+\sqrt{\beta }x),\ \ \ \ \ \ \ \ \sqrt{\beta }x=2(w-\frac{1}{2%
}).  \label{UNO}
\end{equation}%
Therefore, all (asymmetric and symmetric) $q$-Gaussian distribution with $%
-\infty <q<1$ are related to a Beta variable by a shifting of their
arguments.

For the $q$-Gaussian defined by Eq. (\ref{PqMoreAsimetrica}) [or Eq. (\ref%
{AsymetricaMore})], where $1<q<3,$ the change of variables is%
\begin{equation}
w=\frac{1}{2}\Big{(}1+\frac{\sqrt{\beta }x}{\sqrt{1+\beta x^{2}}}\Big{)},\ \
\ \ \ \ \ \sqrt{\beta }x=\frac{(w-\frac{1}{2})}{\sqrt{w(1-w)}}.
\end{equation}

For Eq. (\ref{DistortedExpQless}), the relations are%
\begin{equation}
w=1-\sqrt{\beta }x,\ \ \ \ \ \ \ \ \ \ \sqrt{\beta }x=1-w,  \label{DOS}
\end{equation}%
that is, the modified (and standard) $q$-exponential densities in the
interval $-\infty <q<1$ arise from an axe inversion of a Beta distribution.
Finally, in the interval $1<q<\infty ,$ Eq. (\ref{DistortedUp}), the
transformations are%
\begin{equation}
w=\frac{1}{1+\sqrt{\beta }x},\ \ \ \ \ \ \ \ \ \sqrt{\beta }x=\frac{1-w}{w}.
\end{equation}

The previous relations can be enlighten by using that a variable $W$ obeying
the Beta statistics (\ref{Beta}) can also be written in terms of two
independent Gamma variables $(Y_{1}$ and $Y_{2})$ \cite%
{feller,johnson,kleiber} [Eq. (\ref{jointGamma})]%
\begin{equation}
W=\frac{Y_{1}}{Y_{1}+Y_{2}},\ \ \ \ \ \ \ W^{\prime }=\frac{Y_{2}}{%
Y_{1}+Y_{2}},
\end{equation}%
where the additional variable $W^{\prime }$ is also Beta distributed. In
fact, $W+W^{\prime }=1.$ After a straightforward manipulation, the random
variables associated to the $q$-Gaussian distributions, Eqs. (\ref%
{Aritmetico}) and (\ref{Geometrico}), can\ respectively be rewritten as%
\begin{equation}
\sqrt{\beta }X=(W-W^{\prime }),\ \ \ \ \ \ \ \sqrt{\beta }X=\frac{%
(W-W^{\prime })}{2\sqrt{WW^{\prime }}}.  \label{GaussBetales}
\end{equation}%
while for the $q$-exponentials densities, Eqs. (\ref{XExpoLess}) and (\ref%
{Cociente}), respectively it follows%
\begin{equation}
\sqrt{\beta }X=W^{\prime },\ \ \ \ \ \ \ \ \ \ \ \sqrt{\beta }X=\frac{%
W^{\prime }}{W}.  \label{ExponentialBetales}
\end{equation}%
Both Eq. (\ref{GaussBetales}) and Eq. (\ref{ExponentialBetales}) show the
stretched relation between all the generalized $q$-distributions and Beta
random variables. In fact, any stochastic variable defined by a function
satisfying the symmetry (\ref{symetria}) can be related by a transformation
of variables with a Beta distribution, Eq. (\ref{Beta}).

One may also take the inverse point of view and to explore if the previous
densities can be obtained from Tsallis entropy under a more general
constraint. In fact, any of the extended distributions can be rewritten as $%
P(x)=(\sqrt{\beta }/\mathcal{N}_{q})\exp _{q}[-\beta V(x)].$ This structure
emerges from Tsallis entropy by using a constraint based on a generalized
mean value of $V(x)$ \cite{TsallisBook}. Nevertheless, here the resulting
functions $V(x)$ depend on the parameter $q$ and also on the asymmetry and
distortion factors. Therefore, a relation between Tsallis entropy and the
asymmetric and modified distributions cannot be established in this way.

\subsection{q-triplet for probability densities}

In the context of non-extensive thermodynamics three different values of the
complexity parameter, named as $q$-triplet, are associated to different
physical properties such as the statistics of metastable or quasi-stationary
states, sensitivity to initial conditions, and time-decay of observable
correlations \cite{constantino,burlaga}. Here, we show that three different
values of $q$ allow to indexing the symmetric and unmodified probability
densities. We remark that not any direct relation can be postulated between
both triplets, because here it is established for normalizable objects, $%
\int_{-\infty }^{+\infty }P(x)dx=1.$

We denote by $q_{<1}^{g}$ and $q_{>1}^{g}$\ the complexity indexes of the $q$%
-Gaussian distributions Eqs. (\ref{qGaussMenorUno}) and (\ref{qGaussMayorUno}%
) respectively. Furthermore, $q_{<1}^{e}$ and $q_{>1}^{e}$\ denote the
indexes of the $q$-exponentials, Eqs. (\ref{qExpoMenorUno}) and (\ref%
{qExpoMayorUno}) respectively. The four indexes are given by 
\begin{subequations}
\begin{eqnarray}
q_{<1}^{g} &=&1-\frac{1}{\alpha -1},\ \ \ \ \ \ \ \ \ q_{>1}^{g}=1+\frac{1}{%
\alpha +1/2}, \\
q_{<1}^{e} &=&1-\frac{1}{\alpha -1},\ \ \ \ \ \ \ \ \ q_{>1}^{e}=1+\frac{1}{%
\alpha +1}.
\end{eqnarray}%
We notice that $q_{<1}^{g}=q_{<1}^{e}$ [see Eqs. (\ref{qGaussMenorUno}) and (%
\ref{qExpoMenorUno})]. This equality is expectable because $q$-Gaussian and $%
q$-exponential distributions for $-\infty <q<1$ are related by a linear
change of variables [see Eqs. (\ref{UNO}) and (\ref{DOS})] with a Beta
distribution. Therefore, the complete family of analyzed distributions can
be indexed with only three $q$-parameters: $%
(q_{<1}^{e},q_{>1}^{g},q_{>1}^{e}).$ The previous expressions are equivalent
to 
\end{subequations}
\begin{subequations}
\label{Triplote}
\begin{eqnarray}
\frac{1}{1-q_{<1}^{g}} &=&\alpha -1,\ \ \ \ \ \ \ \ \ \frac{1}{q_{>1}^{g}-1}%
=\alpha +\frac{1}{2}, \\
\frac{1}{1-q_{<1}^{e}} &=&\alpha -1,\ \ \ \ \ \ \ \ \ \frac{1}{q_{>1}^{e}-1}%
=\alpha +1.
\end{eqnarray}%
From here we realize that there exist simple relations between any of the $q$%
-triplet parameters.

\subsection{Applications of the extended distributions}

The assumption\ (\ref{efe})\ lead us with a broad class of probability
densities, which in turn cover the most used probabilities densities arising
from Tsallis entropy maximization. Hence, besides it theoretical interest,
we ask about the possible applications of the asymmetric and modified
distributions.

From the previous analysis, we arrived to the conclusion that asymmetric $q$%
-Gaussian and modified $q$-exponential distributions in the interval $%
-\infty <q<1$ [Eqs. (\ref{AsymetricaLess}) and (\ref{DistortedExpQless})]
are related by a linear change of variables with a Beta distribution.
Therefore, these functions fall in the wide range of applicability of this
distribution \cite{feller,johnson,kleiber}. For example, (unnormalized) Beta
distributions emerge in the statistical description of quark matter (see Eq.
(102) in Ref. \cite{biro}).

The modified $q$-exponential function Eq. (\ref{DistortedUp}) was used in
the description of stock trading volume flow in financial markets \cite%
{queiros,souza,cortines} (named as generalized $q$-Gamma probability
density). This distribution is also known as a Pearson\ Type VI distribution
or alternatively Beta-prime distribution \cite{johnson} (see also \cite%
{kleiber}).

To our knowledge, asymmetric $q$-Gaussian distributions Eq. (\ref%
{AsymetricaMore}) were not used previously. In the present approach, the
asymmetry of these probability densities has a clear dynamical origin. In
fact, associating a cascade process to each Gamma variable, asymmetries
arise whenever the cascades have a different shape index $(\alpha $ and $%
\alpha ^{\prime }).$ Below we discuss the application of these kind of
distributions as a fitting tool in the context of financial signals \cite%
{ausloos} and movement of defects in fluid flows \cite{beck}.

\subsubsection{Log-returns signals on large time windows}

From the price signal $y(t)$ in a financial market it is possible to define
the stochastic process $\tilde{y}(t)=\ln [y(t+\Delta t)/y(t)],$ where $%
\Delta t$ is a constant time interval. This signal gives a simple way of
representing returns in the market. Usually it is studied the \textit{%
normalized log-returns} $Z_{\Delta t}(t)=\ln [\tilde{y}(t)-\langle \tilde{y}%
(t)\rangle ]/\sigma _{\Delta t},$ where $\langle \tilde{y}(t)\rangle $ is
the average and $\sigma _{\Delta t}$\ gives the standard derivation of $%
\tilde{y}(t)$\ for a given $\Delta t.$ Daily closing price values of the
S\&P index for a period of twenty years were analyzed by Ausloos and Ivanova
in Ref. \cite{ausloos}. Assuming a stationary signal, $Z_{\Delta
t}(t)\rightarrow Z_{\Delta t},$ for \textit{large time windows }$(\Delta
t\geq 1$ day), the authors fitted the experimental data with a $q$-Gaussian
like distribution 
\end{subequations}
\begin{equation}
p(z_{\Delta t})=\frac{\sqrt{\beta _{q}}}{\mathcal{N}_{q}}\Big{[}1+\Big{(}%
\sqrt{\beta _{q}}|z_{\Delta t}|\Big{)}^{2\tilde{\alpha}}\Big{]}^{-\frac{1}{%
q-1}},  \label{Financial}
\end{equation}%
where $(1/\mathcal{N}_{q})=\tilde{\alpha}\Gamma (\frac{1}{q-1})/[\Gamma (%
\frac{1}{q-1}-\frac{1}{2\tilde{\alpha}})\Gamma (\frac{1}{2\tilde{\alpha}})]$
and $\sqrt{\beta _{q}}$ depends on the parameters $q$ and $\tilde{\alpha}$\
(see Eqs. (4) and (5) in \cite{ausloos}). This distribution can be obtained
from a superstatistical model assuming, for example, that Brownian particles
diffuse in a potential $U(x)=C|x|^{2\tilde{\alpha}}$ \cite{superstatistics}.
On the other hand, Eq. (\ref{Financial}) can also be recovered from the
present approach based on random Poisson variables. In fact, it follows by
extending symmetrically $(x\rightarrow |x|)$ the stretched $q$-exponential
distribution (\ref{StretchedPower}), with $\alpha ^{\prime }=1/\nu ,$ and
the following replacements $\tilde{x}\rightarrow |z_{\Delta t}|,$ $\tilde{%
\beta}\rightarrow \beta _{q},$ $\nu \rightarrow 2\tilde{\alpha},$ and $%
\alpha \rightarrow (\frac{1}{q-1}-\frac{1}{2\tilde{\alpha}}).$ Random
numbers generation is achieved by introducing an extra stochastic variable
that with probability one half defines their sign (positive or negative). 
\begin{figure}[tbp]
\includegraphics[bb=22 100 570 1010,width=8.75cm]{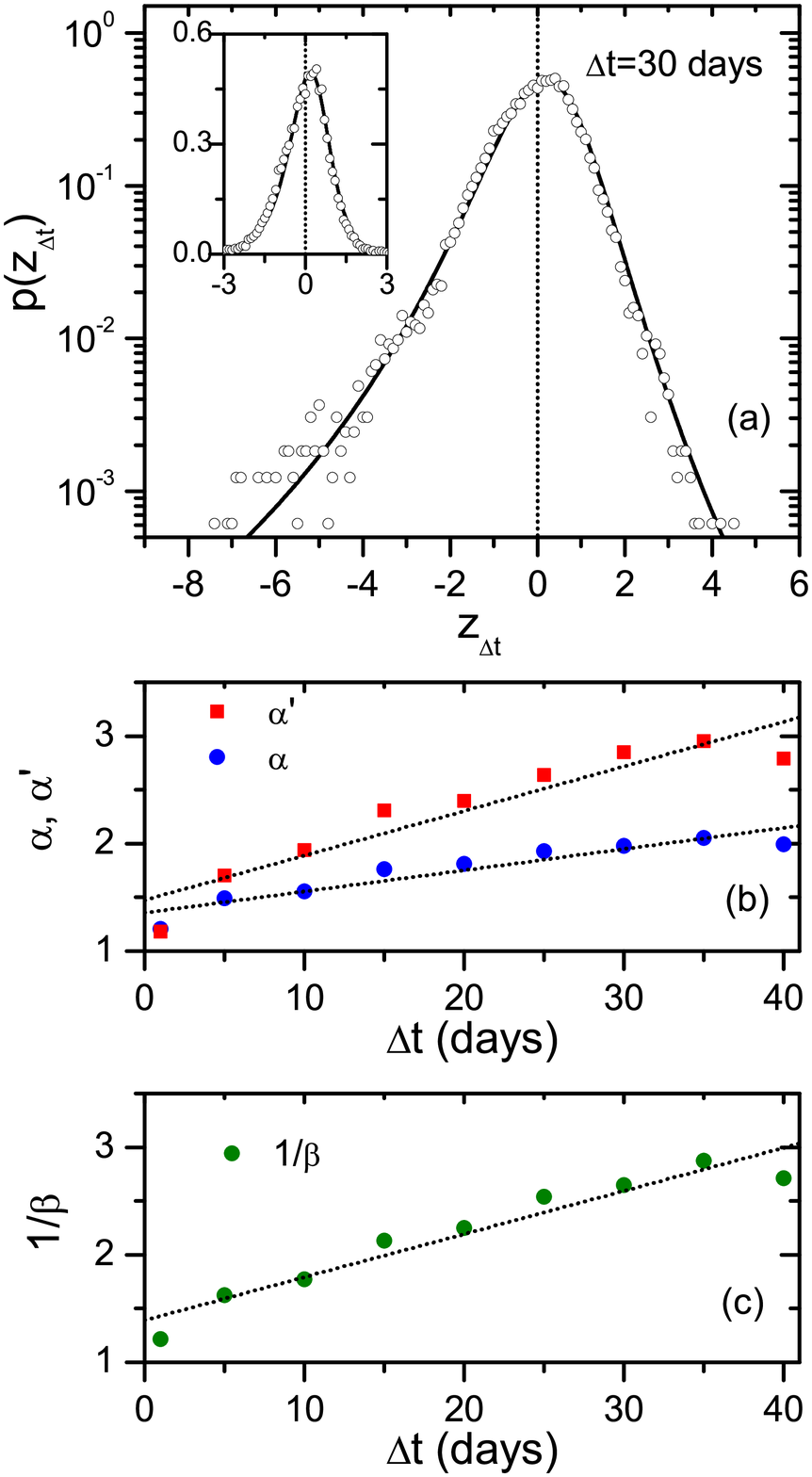}
\caption{(a) Probability density $p(z_{\Delta t})$ of normalized log-returns
for the S\&P index (circles). The full line corresponds to Eq. (\protect\ref%
{PqMoreAsimetrica}) (see text). The inset shows the peak region. The
parameters are $\protect\alpha =1.97,$ $\protect\alpha ^{\prime }=2.85,$ and 
$\protect\beta =0.37.$ From (\protect\ref{qGaussMayorUno}) it follows $%
q=1.34 $ and $a=-0.87.$ (b) Dependence with $\Delta t$ of the shape
parameters $\protect\alpha $ and $\protect\alpha ^{\prime }.$ Dotted line
linear fit. (c) Parameter $\protect\beta .$}
\end{figure}

Eq. (\ref{Financial}) develops an asymptotic $(\sqrt{\beta _{q}}|z_{\Delta
t}|\gg 1)$ power-law behavior. Nevertheless, the authors also find that the
experimental data are not consistent with the symmetry $p(z_{\Delta
t})=p(-z_{\Delta t}).$ In particular, the power-law behaviors have different
exponents for positive and negative values. Similar \textit{asymmetries}
were found previously in Ref. \cite{stanley}.

Here, we may associate the observed asymmetry of the data to two cascades
mechanisms, each one being represented by a Gamma random variable. In a
roughly way, the difference between both variables can be associated to
different networks properties related to the propagation of information that
support an increased or decreased future value. The complete system response
is defined by Eq. (\ref{Geometrico}), that is a geometric mean value of the
driving fluctuations. Hence, instead of using Eq. (\ref{Financial}), we
propose to fit the probability density of the log-returns with the
asymmetric $q$-Gaussian distribution (\ref{AsymetricaMore}) under the
shifting $P(x)\rightarrow P(z_{\Delta t}+\langle X\rangle ),$ where $\langle
X\rangle $ is given by Eq. (\ref{X}).

In order to check this proposal, here we analyze the daily closing price
values of the S\&P index \cite{yahoo} for the period between Jan. 3, 1950
and\ Dec. 3, 2014, which provides a 16336 data base. In Fig. 5(a) we show
the \textquotedblleft experimental\textquotedblright\ probability
distribution (circles) for $\Delta t=30$ days. The data are clearly
asymmetric, which in fact are fitted by Eq. (\ref{AsymetricaMore}) (full
line). Its characteristic parameters $\alpha ,$ $\alpha ^{\prime },$ and$\
\beta $ were determinate by minimizing the global error. Based on the
quadratic global error $\sum (p_{\exp }-p_{theory})^{2}/p_{\exp },$ we
checked that for a wide range of $\Delta t$ the asymmetric distribution
provides a better fitting than Eq. (\ref{Financial}).

In\ Fig. 5(b) we plot $\alpha $ and $\alpha ^{\prime }$ as function of $%
\Delta t.$ In the limit $\Delta t\rightarrow 0$ the asymmetry vanishes $%
(\alpha \simeq \alpha ^{\prime }).$ Furthermore, in the plotted interval,
both shape parameters increase linearly with $\Delta t.$ For higher values
of $\Delta t$ an irregular-logarithmic-like grow\ behavior is observed (not
shown). For $\Delta t\gtrsim 500$ the distributions approaches normal
Gaussian ones. This limit is consistent with the grow of $\alpha $ and $%
\alpha ^{\prime }$ [see Eq. (\ref{qGaussMayorUno})]. On the other hand, we
find that $1/\beta $ also increases linearly with $\Delta t,$ Fig. 5(c).
This (diffusive) behavior is also found for intervals $\Delta t$ minor than
a day \cite{fuentes}.

Extra analysis and ingredients are necessary for explaining the linear
behaviors shown in Figs. 5(b) and (c). On the other hand, Fig. 5(a) shows
that the proposed probability density provides a reasonable and alternative
fitting to that based on Eq. (\ref{Financial}), which in turn is able to
capture the observed asymmetries.

\subsubsection{Defect velocities in inclined layer convection}

In Ref. \cite{beck} Daniels, Beck, and Bodenshatz studied the motion of
point defects in thermal convection patterns in an inclined fluid layer
(heated from below and cooled from above), a variant of Rayleigh-B\'{e}nard
convection. Due to the inclination the system is anisotropic. The
(experimental) probability distribution of the (positive and negative)
defect velocities is different in the transverse ($\hat{x},$ across rolls)
and longitudinal ($\hat{y},$ along rolls, uphill-downhill) directions.\ In
the transverse direction the velocity $(v_{x})$ can be fit with a symmetric $%
q$-Gaussian distribution $(q\simeq 1.4).$ Nevertheless, in the longitudinal
direction $(v_{x})$ the distribution, depending on a dimensionless
temperature $\varepsilon $ (see details in \cite{beck}), develops strong
asymmetries. In that situation, Tsallis distributions, even defined with a
cubic potential, are unable to fit the experimental data (see Fig. 4(a) in 
\cite{beck}). As shown in the next figure, these asymmetries can be fitted
with the probability densities introduced previously. 
\begin{figure}[tbp]
\includegraphics[bb=18 284 420 595,width=8.75cm]{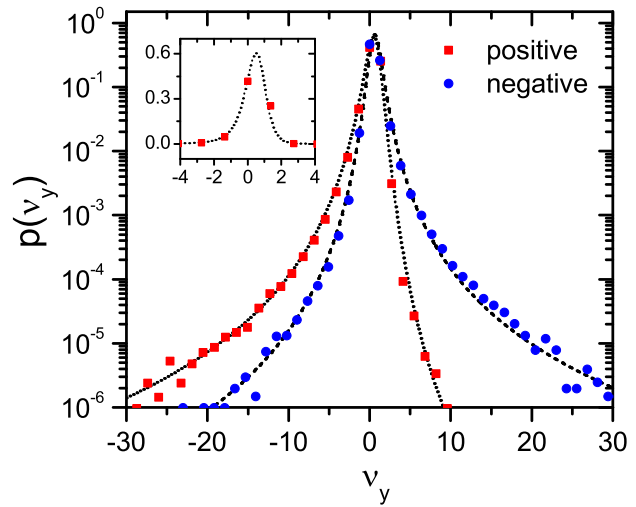}
\caption{Probability density $p(\protect\nu _{y})$ of the normalized
velocity $\protect\nu _{y}=v_{y}/\protect\sqrt{\langle v_{y}^{2}\rangle
-\langle v_{y}\rangle ^{2}}$ for positive and negative defects in inclined
layer convection, $\protect\varepsilon =0.08$ (see \protect\cite{beck}\ for
details). The fitting (dotted and dashed lines) correspond to Eq. (\protect
\ref{AsymetricaMore}) (see text). For positive defects (red squares
experimental data) the parameters are $\Delta =0.62,$ $\protect\alpha =1.60,$
$\protect\alpha ^{\prime }=2.71,$ which lead to $q=1.37$ and asymmetry $%
a=-0.55.$ For negative defects (blue circles experimental data) the
parameters are $\Delta =0.58,$ $\protect\alpha =1.67,$ $\protect\alpha %
^{\prime }=1.38,$ which lead to $q=1.49$ and $a=0.14.$ The inset shows the
peak region for positive defects.}
\end{figure}

In Fig. 6 we show a set of experimental data $(\varepsilon =0.08)$ \cite%
{karen}, corresponding to the probability density $p(\nu _{y})$ of the
dimensionless velocity $\nu _{y}\equiv v_{y}/\sqrt{\langle v_{y}^{2}\rangle
-\langle v_{y}\rangle ^{2}}$ of positive and negative defects (see also Fig.
4(a) in \cite{beck}). We find that these data can be very well fitted with
the distribution Eq. (\ref{AsymetricaMore}) under the associations $%
x\rightarrow \nu _{y}-\Delta /\sqrt{\beta }=(v_{y}/\sqrt{\langle
v_{y}^{2}\rangle -\langle v_{y}\rangle ^{2}})-\Delta /\sqrt{\beta }.$ Hence, 
$p(\nu _{y})=\sigma P(\sigma \nu _{y}-\Delta /\sqrt{\beta }),$ where $\sigma
\equiv \sqrt{\langle X^{2}\rangle -\langle X\rangle ^{2}},$ follows from
Eqs. (\ref{X}) and (\ref{X^2}). Furthermore, we introduced an extra
dimensionless parameter $\Delta $ that only shifts the complete
distribution. Due to the previous rescaling, $p(\nu _{y})$ does not depend
on the parameter $\beta .$ The parameters $\alpha ,$ $\alpha ^{\prime },$
and $\Delta $ were determinate such that the global error is minimized. Even
when the asymmetry is appreciable, the maximum of the distribution [Eq. (\ref%
{xm})] is around the origin. In fact, the influence of the shift introduced
by $\Delta $ is only appreciable around the origin (note that in both cases $%
|\Delta |<1).$ For both positive and negative defects a very well fitting is
obtained. We also checked that a similar fitting is obtained for higher
values of the dimensionless temperature, $\varepsilon =0.17,$ where the
distribution asymmetry is smaller than that shown in Fig. 6 (see Fig. 4(b)
in \cite{beck}). Hence, we conclude that dynamics of the defects velocity
may be though as being governed by two cascade mechanisms with different
statistical properties, such as that defined by Eq. (\ref{Geometrico}).
While a rigorous derivation of this interpretation is not developed here,
the quality of the fitting gives a consistent support to the proposed
theoretical frame.

\section{Conclusions}

The present approach relies on expressing the variable of interest,
associated to a given complex system, as a function of two independent Gamma
random variables. These variables represent intrinsic fluctuations that
drive the system. In addition, the complexity of the dynamics is represented
by a random-system-response that also depends, in a non-linear way, on the
fluctuations. Writing the system response in terms of a generalized mean
value, Eq. (\ref{GenerlizedMoment}), we showed that the arithmetic and
geometric cases allow us to introduce a class of asymmetric $q$-Gaussian
distributions. In the symmetric case, for any value of the complexity
parameter $q,$ they recover densities that follow from Tsallis entropy
maximization. A similar approach applies for $q$-exponential distributions,
which become defined in terms of a distortion parameter. We also showed that
the complete family of obtained distributions can be related via a change of
variables with a Beta distribution. A $q$-triplet was derived for the
symmetric and unmodified distributions.

On one side, these results define an alternative numerical tool for random
number generation obeying the previous statistical behaviors. On the other
hand, the present approach may provide an alternative and very simple basis
for understanding statistical behaviors in complex dynamics. Of special
interest is the possibility of relating any asymmetry in the probability
distributions with different underlying cascade mechanisms. We have shown
that in fact asymmetric Poissonian $q$-Gaussian densities $(1<q<3)$ provide
a very well fitting to the statistical distribution of log-return signals in
financial markets (Fig. 5) as well as the probability distribution of the
velocity of moving defects in inclined layer convection \cite{beck} (Fig.
6). Therefore, the derivation of the present approach from deeper
microscopic or mesoscopic descriptions is an issue that with certainty
deserves extra analysis. The possibility of recovering asymmetric
distributions or the Beta statistics from non-extensive thermodynamics also
remains as an open problem.

\section*{Acknowledgment}

I am indebted to K. E. Daniels for fruitful discussions and for sending the
experimental data of defect turbulence. This work was supported by CONICET,
Argentina.

\end{document}